\documentclass[journal=jctcce,manuscript=article,]{achemso}

\usepackage[version=3]{mhchem} 

\usepackage{longtable,booktabs}

\usepackage{graphicx,grffile}
\makeatletter
\def\maxwidth{\ifdim\Gin@nat@width>\linewidth\linewidth\else\Gin@nat@width\fi}
\def\maxheight{\ifdim\Gin@nat@height>\textheight\textheight\else\Gin@nat@height\fi}
\makeatother
\setkeys{Gin}{width=\maxwidth,height=\maxheight,keepaspectratio}

\IfFileExists{parskip.sty}{%
\usepackage{parskip}
}{
\setlength{\parindent}{0pt}
\setlength{\parskip}{6pt plus 2pt minus 1pt}
}

\providecommand{\tightlist}{%
  \setlength{\itemsep}{0pt}\setlength{\parskip}{0pt}}

\author{Sky(Yixiang) Zhang}

\affiliation[DOC, Tsinghua University]{Department of Chemistry, Tsinghua University, Beijing, 100084, P.R.
China}

\author{Hai Xiao}

\email{haixiao@tsinghua.edu.cn}

\affiliation[DOC, Tsinghua University]{Department of Chemistry, Tsinghua University, Beijing, 100084, P.R.
China}

\author{Jun Li}

\email{junli@tsinghua.edu.cn}

\affiliation[DOC, Tsinghua University]{Department of Chemistry, Tsinghua University, Beijing, 100084, P.R.
China}

\title{Efficient Chain-of-States Approach for Locating Transition State via
Spherical Optimization}


\begin{document}
%
\begin{abstract}
The Chain-of-states(CoS) methods like nudge elastic band(NEB) method can
be used to determine the minimum energy path (MEP) and transition state
(TS) between two end local minima. However, the CoS methods are
inefficient for difficult cases where the two ends are far apart with
chemically insignificant part(s) in the MEP. We present here a method
based on spherical optimization (SOPT), in which the SOPT method
generates model end structures for CoS methods under the constraint of
constant root-mean-square distance (RMSD) between two ends that is
chosen to cover only the chemically significant part. We demonstrate the
robustness and efficiency of our method with two examples, the CHOH
dissociation channel and the first step of Aldol reaction. In both
cases, the SOPT-based NEB calculations always reach the convergence to
the correct MEPs with much less computational cost, whereas the regular
NEB calculations fail under certain setups.
\end{abstract}

\section{INTRODUCTION}

Finding the minimum energy path (MEP) between two local minima (the
initial and final states) and thus locating the transition state (TS) is
a key topic in theoretical and computational chemistry. MEP is the
highest statistically weighted path connecting the two states, and the
maximum along the path is identified as the TS, which is a first-order
saddle point on the potential energy surface (PES). Once the MEP and TS
are determined, the transition rate between the two states can be then
estimated using the transition state theory.

Various methods have been proposed to determine the MEP, and they can be
generally classified into two categories: the single-ended methods and
the double-ended methods, with the latter also known as the
chain-of-states (CoS) methods. In applying the CoS methods such as the
nudged elastic band (NEB) method
\cite{Henkelman2000Improved, NEB-book, NEB2012, Sheppard2008Optimization, Sheppard2011Paths, Sheppard2012A},
two end states are required beforehand, and an initial guess of states
composing the chain is generated (typically by linear interpolation
between the two end states, or the image dependent pair potential (IDPP)
method \cite{IDPP}). Optimization routine is then executed until one
image of the chain reaches the TS point. Therefore, the robustness and
efficiency of the optimization algorithm and the quality of initial
guess for the chain both dictate the speed of transition state search.
NEB (including its variations like CI-NEB \cite{CINEB}), as one of the
best CoS methods, is frequently used to find MEP because of its
robustness and efficiency.

Concerning the initial guess of the chain, the external degrees of
freedom like translation and rotation may slow down or even prevent the
convergence to MEP. A method based on quaternion algebra solved the
problem by minimizing the root-mean-square deviation (RMSD) of atomic
positions between two images \cite{dof-quaternion} . However, there are
cases where one of the ends cannot be optimized, or the optimized
geometry differs drastically from the other one. For example, the
formaldehyde dissociation reaction (CHOH \(\rightarrow\) CO + \ce{H2})
has one end composed of two molecules barely bound by very weak
interaction, which leads to a flat PES that poses a great difficulty for
the geometry optimization to reach convergence. Even if it converges
with loose criteria, the resulted structure is likely far too different
from the other end. This means that more images and iterations are
needed for MEP convergence, although the part of MEP near this one end
is essentially diffusion and is of little interest. Generally, for any
MEP, the part near one or both ends is dominated by diffusion and thus
is not interesting, but the images are still needed for converging the
chain, and these extra images slow down the convergence significantly.

One solution is to use an artificial model structure as the end instead
of a local minimum. This implies the necessity of geometry optimization
for an end structure with constraints, and thus inspires us to compose a
properly formulated constraint for optimizing the model end structure
without increasing the distance between two ends. In the present
article, we use the spherical optimization(SOPT) method
\cite{Abashkin1994Transition} to achieve constrained optimization of
model end structure(s) with fixing the RMSD between two ends. This
method enables faster MEP convergence with less images needed. Although
NEB is used as the MEP searching algorithm here for simplicity, this
method is applicable to any CoS methods for pre-optimizing model end
structures.

\section{METHODS}

\subsection{NEB Method and Minimize RMSD with Quaternion Algebra}

NEB method is one of the most popular MEP searching methods. In
practically applying the NEB method, a discrete representation of the
MEP is generated initially, the elements of which are referred as
``images'', {[}\(R_0, R_1, ..., R_N\){]}. The NEB force on image \(i\)
is modified as

\begin{align}
\begin{split}
F_i &= F_i^{\parallel} + F_i^{\perp}\\
F_i^{\perp}     &= - ( \nabla E(R_i) - \nabla E(R_i) \cdot \hat{\tau} \hat{\tau}) \\
F_i^{\parallel} &= k (|R_{i+1} - R_i| - |R_i - R_{i-1}|) \hat{\tau}\\
\end{split}
\end{align}

where \(k\) is the coefficient of stiffness and is usually 0.1 eV/\AA{},
and \(\hat{\tau}\) is the local unit tangent.A simple estimation can be

\begin{align}
\hat{\tau} = \frac { R_{i+1} - R_{i-1}} {|R_{i+1} - R_{i-1}|}
\end{align}

Note that the external degrees of freedom including translation and
rotation introduce inefficiency into the MEP searching, because part of
the forces acting on the images contributes to translating and rotating
the images that are chemically insignificant. This problem is solved by
minimizing the RMSD between images with quaternion algebra
\cite{dof-quaternion} . The square of RMSD, called residual, is defined
as

\begin{align}
\mathcal{E} = \frac 1 N \sum_{k=1}^N|\mathcal{R}x_k' + g - y_k'|^2
\end{align}

where \(N\) is the number of atoms of the system, \(x_\text{k}'\),
\(y_\text{k}'\) are the coordinates of atom k in two images, \(g\) and
\(\mathcal{R}\) are translation vector and rotation matrix,
respectively. \(g\) is simply defined as the difference between centers
of mass of two images, and \(\mathcal{R}\) is calculated with quaternion
algebra \cite{dof-quaternion} and the details of the algorithm are
skipped here.

Additionally, as elaborated in the previous section, it is much more
efficient to introduce model end structures for NEB that allow a short
path with inclusion of only key images and thus enable efficient
convergence. Thus, an optimization technique is required for relaxing
the internal forces without increasing the distance between two ends.

\subsection{Spherical Optimization}

Consider first that one of the ends employs the model structure (noted
as \(x\), and the other end as \(y\); the system is with \(N\) atoms),
with the constraint of keeping the RMSD between two ends fixed, it is
converted to a constrained optimization problem, which leads to
esentially the SOPT method by Y. Abashkin and N. Russo
\cite{Abashkin1994Transition},

\begin{align}
\begin{split}
&\min_{\vec{x}} E   =    E(x_1, x_2, ..., x_{n})\\
&\text{subject to}\quad (x_1 - y_1)^2 + (x_2 - y_2)^2 + ... + (x_{n}-y_{n})^2 = R^2
\end{split}
\end{align}

where \(R\) is the initial RMSD, and \(n=3N\) is the total degrees of
freedom. To solve this optimization problem, the constraint function is
rewritten by choosing an index \(q\) so that

\begin{align}
x_q = f(x_1, x_2, ..., x_{n-1}, R) = y_q \pm \sqrt{R^2 - \sum_{i\ne q}^{n}(x_i - y_i)^2}
\label{eq:x_q}
\end{align}

For numerical stability, \(q\) is chosen so that \(|x_q - y_q|\) is
maximized, and the sign in Eq. \ref{eq:x_q} is carefully chosen so that
it matches the reality. For simplicity, we swap \(q\) with \(n\), and
the energy function is rewritten by including the constraint,

\begin{align}
E' = E(x_1, x_2, ..., x_{n-1}, f(x_1, x_2, ..., x_{n-1}, R))
\end{align}

The derivative of \(x_n\) with respect to \(x_i\) is

\begin{align}
\begin{split}
\frac {\partial x_n}{\partial x_i} =& \frac {\partial f}{\partial x_i} = -\frac{x_i-y_i}{x_n-y_n}
\end{split}
\end{align}

and the force is rewritten as

\begin{align}
\begin{split}
F'_i =& -\frac{\partial E'}{\partial x_i} = -(\frac{\partial E}{\partial x_i} + \frac{\partial E}{\partial x_{n}} \frac{\partial x_n}{\partial x_i})\\
     =& F_i - F_n \frac{x_i-y_i}{x_n-y_n}
\end{split}
\label{eq:spherical-force}
\end{align}

where \(F\) is the force obtained from electronic structure calculation,
and \(F'\) is a \(n-1\) vector. Thus, we convert this particular
constrained optimization to a normal optimization problem with \(n-1\)
variables. Regular optimization techniques can be used for solving the
problem, such as steepest descent, conjugate gradient, BFGS, and L-BFGS.
After the optimization, the internal force is relaxed while keeping the
RMSD between two ends constant, thus speeding up further NEB
calculations, as we will see in the examples.

\subsection{Flowchart of Spherical Optimization}

We provide the flowchart for the algorithm of spherical optimization
below,

\begin{enumerate}
\def\labelenumi{\arabic{enumi}.}
\tightlist
\item
  One end is assigned as the initial state that will be optimized, and
  the other is assigned as the partner.
\item
  Use the quaternion method mentioned above to minimize the RMSD between
  two ends.
\item
  Calculate the force acting on each atom in the initial state.
\item
  Recalculate the modified force with Eq. \ref{eq:spherical-force}.
\item
  Perform optimization with a chosen technique.
\item
  Repeat 3-5 until the forces meet the criteria.
\end{enumerate}

If both ends need to be optimized, then swap the initial and final
states, and repeat the procedure again. In principle, both ends should
be optimized repeatedly, but it is found that the result is good enough
for NEB calculation with optimizing each end only once.

\section{RESULTS}

In the following, NEB with SOPT (NEB-SOPT) is applied to two examples,
the CHOH dissociation channel and the first step of Aldol reaction.
Regular NEB (NEB-R) calculations are performed as well for comparison.
Both NEB-SOPT and NEB-R are performed with the climbing image option,
and the linear interpolation over Cartesian coordinates is used for
initial NEB chain generation. Gaussian 09 \cite{g09} is used as the
calculator of electronic structures and forces, as well as for geometry
optimizations without constraint. The B3LYP functional and 6-31G(d)
basis sets with default parameters in Gaussian 09 are used in all the
cases if not specified. The BFGS function provided in ASE
\cite{ase, ase-old} is used as geometry optimizer in SOPT and NEB
calculations with step size of 0.2\AA{} and the maximum number of steps
of 120. The convergence max forces are all set to be 0.2eV/\AA{}, while
it is worth mentioning that the max force could be set to 0.5eV/\AA{} in
SOPT to reduce the iteration number and the result is still sufficiently
good.

\subsection{CHOH dissociation channel}

\begin{figure}
\centering
\includegraphics{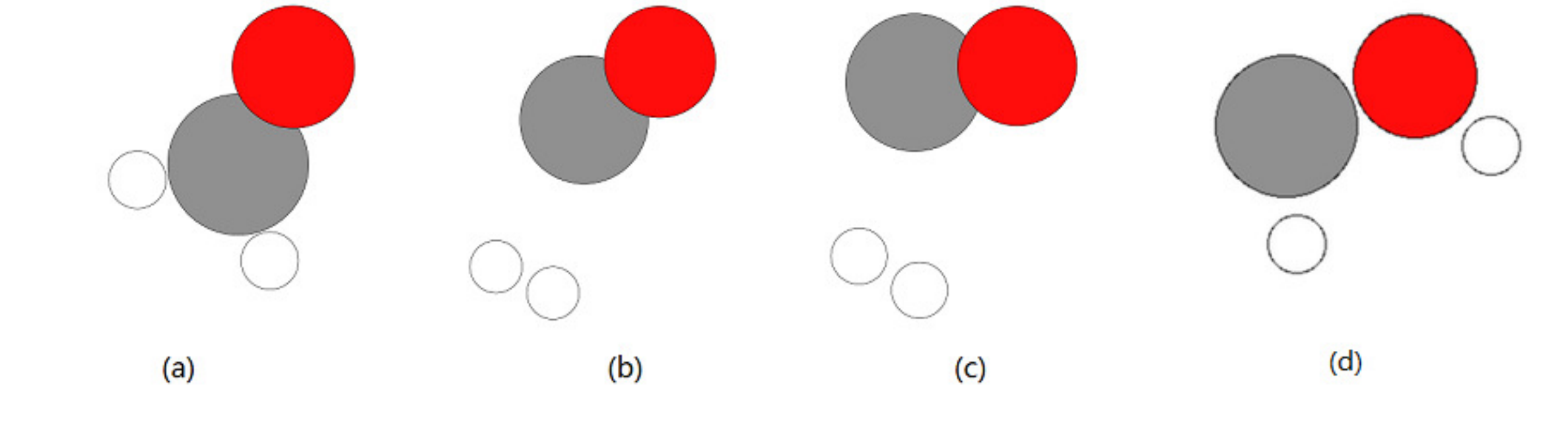}
\caption{(a) The CHOH molecule, (b) CO+\ce{H2} by regular optimization,
(c)CO+\ce{H2} by SOPT, and (d) TS given by NEB3-R, which is wrong.
\label{fig:choh_dc_ends}}
\end{figure}

CHOH MEP searching has been taken as one of the benchmarks for MEP and
TS searching methods. In this study, the dissociation channel of CHOH
(CHOH-dc) is selected as our test case, in which the CHOH molecule is
dissociated into CO and \ce{H2}. In applying the NEB-R method, a
structure composed of CO and \ce{H2} should be prepared and optimized to
local minimum. However, since there is only weak interaction between CO
and \ce{H2}, the structure cannot be optimized with routine convergence
criteria. So we start from an artificially modeled structure and
optimize it with BFGS provided in ASE, and the max force is set as 0.2
eV/\AA{}, which is higher than ordinary. The structures by regular
optimization and SOPT are shown in Fig.\ref{fig:choh_dc_ends}(b) and
Fig.\ref{fig:choh_dc_ends}(c), respectivly. The energy difference
between two structures is only 0.04 eV, implying the weak interaction
between two molecules. But the geometry structures are quite difference.
Fig.\ref{fig:choh_dc_ends}(c) shows a much shorter distance between CO
and \ce{H2} than that in Fig.\ref{fig:choh_dc_ends}(b), and the
structure in Fig.\ref{fig:choh_dc_ends}(c) shows a special relative
position between the two molecules that does not exist in
Fig.\ref{fig:choh_dc_ends}(b). This implies that the structure in
Fig.\ref{fig:choh_dc_ends}(c) is optimized to a neighbor of TS,
suggesting the suitability of SOPT for accerlating NEB calculations.

\begin{figure}
\centering
\includegraphics{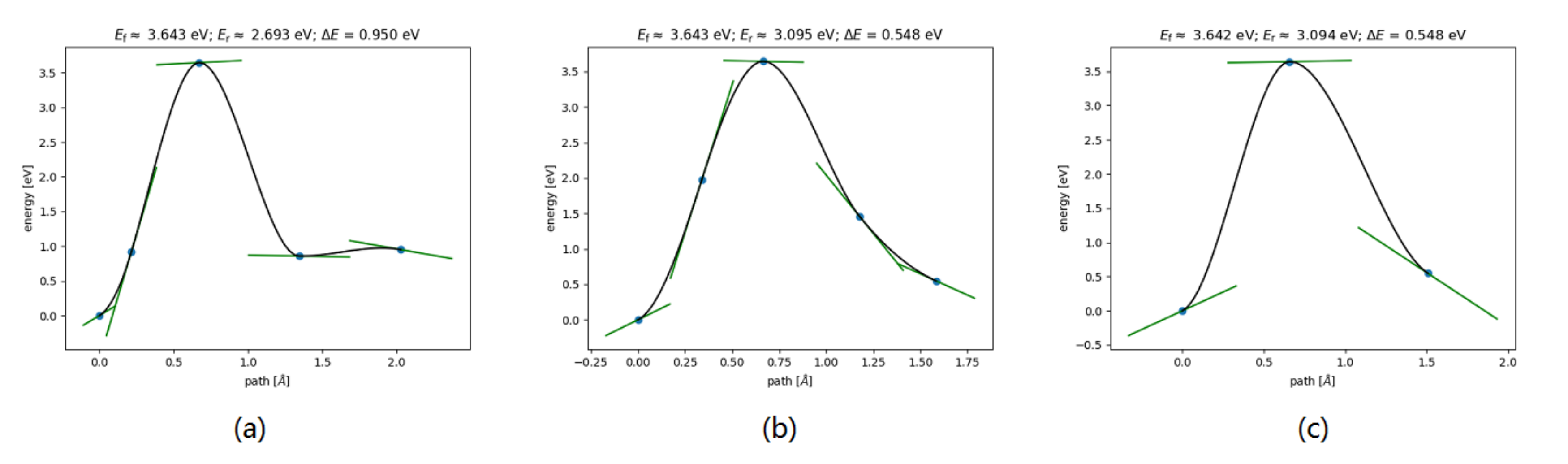}
\caption{The MEPs of CHOH-dc by (a) NEB5-R, (b) NEB5-SOPT, and (c)
NEB3-SOPT. \label{fig:choh_dc_neb}}
\end{figure}

NEB calculations are performed with 5 and 3 images, and noted as NEB5
and NEB3, respectively. The numbers of iterations needed are shown in
Table \ref{table:noiter}. In NEB5, both NEB-SOPT and NEB-R calculations
are converged, but the MEPs given by two methods are different in some
aspects, as shown in Fig.\ref{fig:choh_dc_neb}. The difference in
activation energy is caused by the convergence criteria and is
acceptable. However, in NEB-R calculation shown in
Fig.\ref{fig:choh_dc_neb}(a), image 1 and 2 are very close, and image 4
is a shallow local minimum on MEP curve, whereas in
Fig.\ref{fig:choh_dc_neb}(b), there is no local minimum on the curve
between two ends, with the images on the curve distributed evenly, and
the curve by NEB-SOPT is much smoother than that by NEB-R. The length of
entire reaction path is controlled to no more than 1.75\AA{} in
NEB-SOPT, whereas it goes to almost 2.0\AA{} in NEB-R. The SOPT
introduces pre-optimization of the ends, while such additional
computational cost is negligible. However, the iteration numbers needed
for NEB-R and NEB-SOPT are 28 and 10, respectively, which means that
NEB-SOPT is computationally cheaper by more than 60\% in terms of
iterations and force calculations.

Because of the simplicity of the reaction, we also tried 3 images for
NEB-R and NEB-SOPT, and the comparison is more revealing. For NEB-SOPT,
only 13 iterations are needed for a correct convergence as shown in
Fig.\ref{fig:choh_dc_neb}(c), whereas NEB-R uses 36 iterations, but
converges to a wrong MEP and the TS is shown in
Fig.\ref{fig:choh_dc_ends}(d). Since NEB3 saves 2/3 of force evaluations
than NEB5, NEB-SOPT can practically save 85\% of force calculations in
this case, and the result is still correct. This shows the robustness
and efficiency of the method.

\begin{longtable}[]{@{}lrr@{}}
\caption{Number of iterations used by NEB-R and NEB-SOPT
\label{table:noiter}}\tabularnewline
\toprule
Application & NEB-R & NEB-SOPT\tabularnewline
\midrule
\endfirsthead
\toprule
Application & NEB-R & NEB-SOPT\tabularnewline
\midrule
\endhead
CHOH-dc(NEB5) & 28 & 11\tabularnewline
CHOH-dc(NEB3) & 36(wrong TS) & 13\tabularnewline
Aldol-1(NEB11) & 111 & 33\tabularnewline
Aldol-1(NEB5) & fail & 30\tabularnewline
\bottomrule
\end{longtable}

\begin{figure}
\centering
\includegraphics{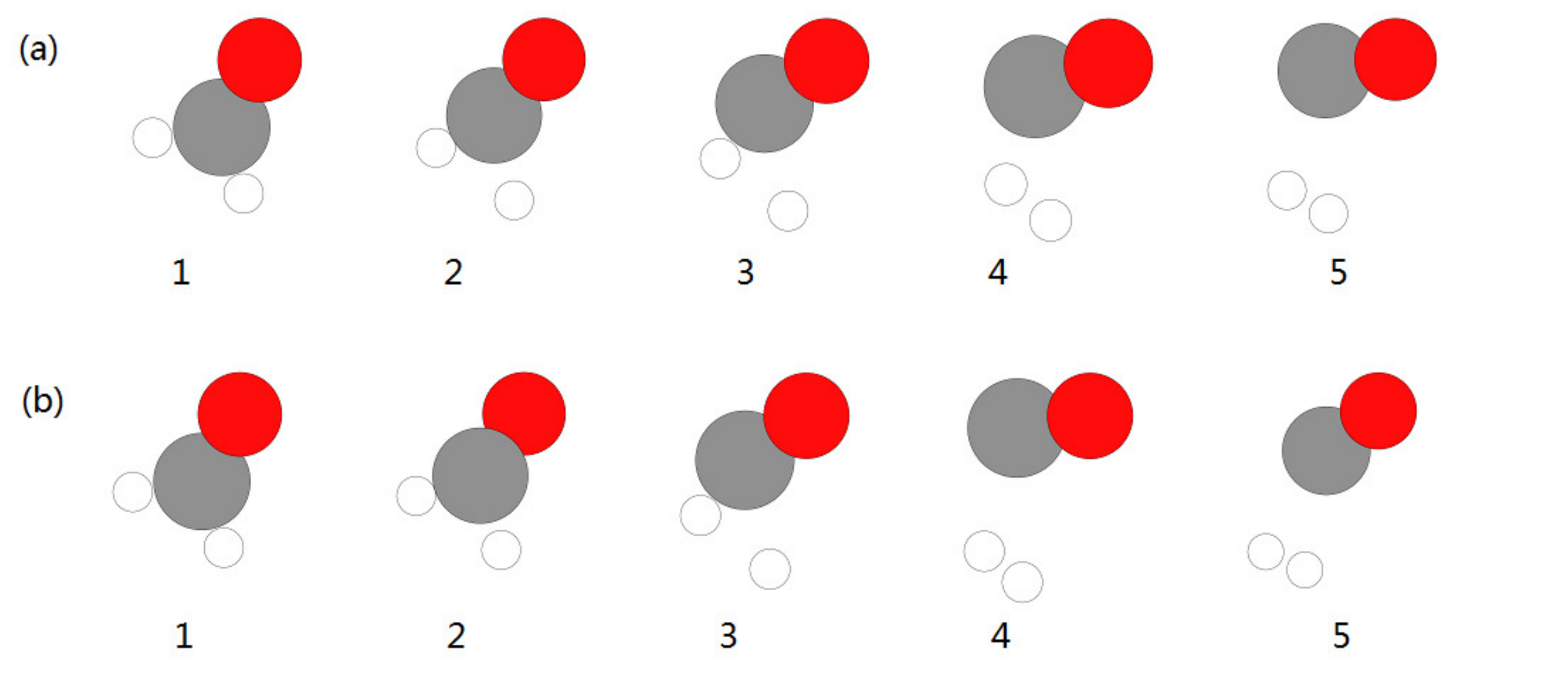}
\caption{The images in MEPs by (a) NEB5-R and (b) NEB5-SOPT.
\label{fig:choh_dc_images}}
\end{figure}

To better understand the SOPT, we illustrate each image of CHOH-dc MEPs
by NEB5-R and NEB5-SOPT in Fig.\ref{fig:choh_dc_images}(a) and
Fig.\ref{fig:choh_dc_images}(b), respectively. It shows that the image 4
in NEB5-R is almost identical to the image 5 in NEB-SOPT. This means
that only 4 of 5 images are relevant in NEB5-R and the last image is
only from the diffusion process that is uninsteresting. In other words,
our method removes insignificant images by SOPT.

\subsection{The First Step of Aldol Reaction}

\begin{figure}
\centering
\includegraphics{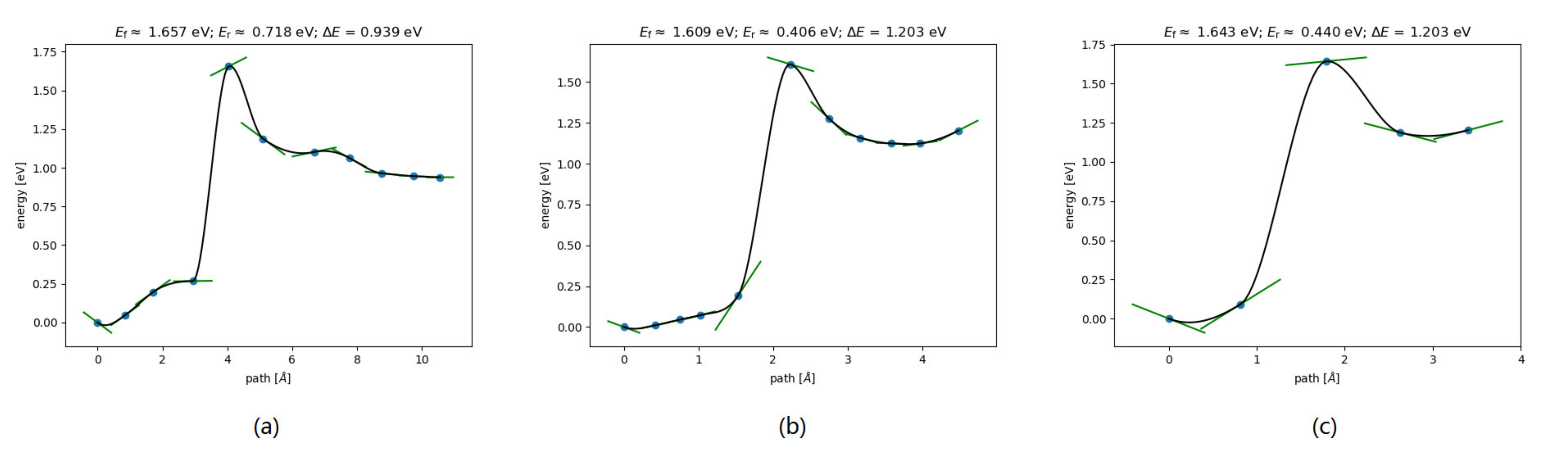}
\caption{The MEPs of Aldol-1 by (a)NEB11-R, (b) NEB11-SOPT, and (c)
NEB5-SOPT. \label{fig:Aldol-1-nebs}}
\end{figure}

The first step of Aldol reaction(Aldol-1) is chosen as a test case, in
which formaldehyde(CHOH) and ethenol (\ce{CH2CHOH}) couple to form
3-hydroxypropionaldehyde(\ce{C3H6O2}). This case has been tested by
Keiji Morokuma et al. with the GRRM program \cite{GRRM}. In this case,
PBE/6-31G(d) is used for SOPT to testify the robustness and generality
of the method, because it's much cheaper than B3LYP, but the calculated
force is relatively correct. This may introduce some error into the
energies of end points because of the optimized geometry difference
between two functionals, but the result is still fine. We first test
NEB-R and NEB-SOPT with 5 images. 30 iterations are needed for NEB5-SOPT
to converge to the MEP shown in Fig.\ref{fig:Aldol-1-nebs}(a), but
NEB5-R cannot converge after 120 iterations, and the intermediate image
structures are broken and unphysical, even with a quite small step size
of 0.05 \AA{}. We further test NEB-R with 7 and 9 images, and NEB-R
still cannot converge. With 11 images, NEB-R finally converges with 111
iterations, whereas NEB-SOPT uses only 33 iterations, as listed in Table
\ref{table:noiter}, and the MEPs are shown in
Fig.\ref{fig:Aldol-1-nebs}(b) and Fig.\ref{fig:Aldol-1-nebs}(c).

\begin{figure}
\centering
\includegraphics{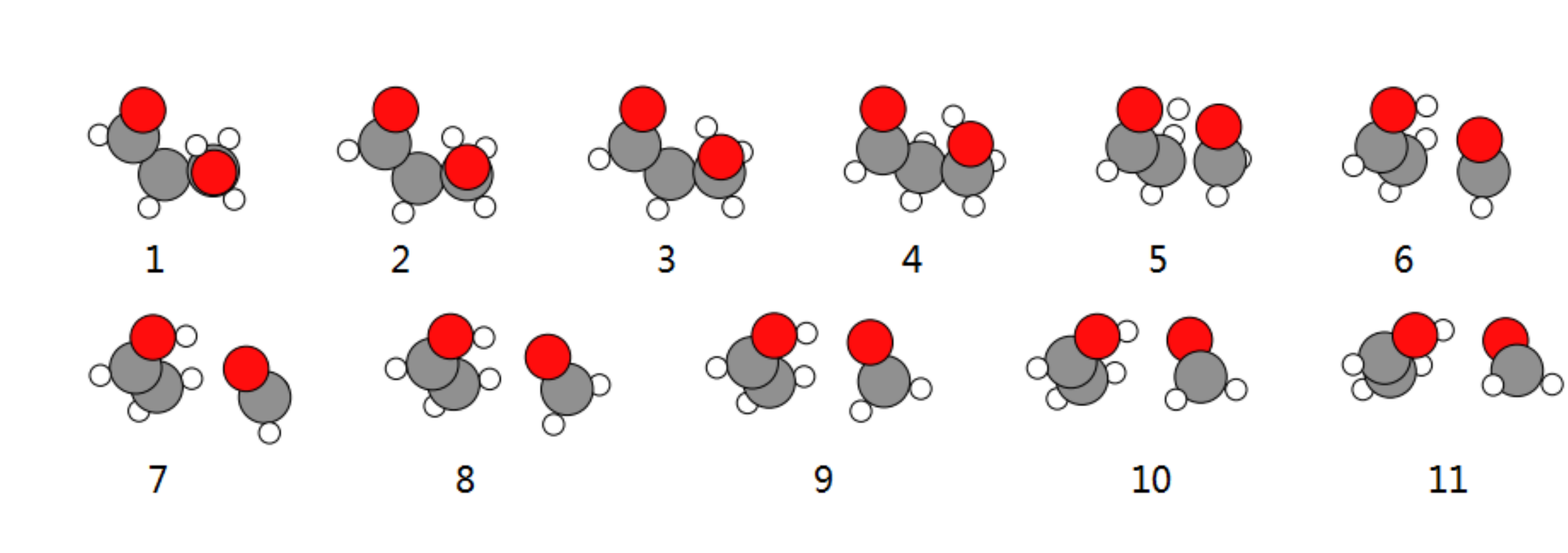}
\caption{The images in the Aldol-1 NEB11-R MEP.
\label{fig:Aldol-1-normal-images}}
\end{figure}

In Fig.\ref{fig:Aldol-1-nebs}(a), the MEP of NEB5-SOPT is quite smooth,
while the image 4 with energy a little lower than the image 5 is the
result of larger \(\mathcal{R}\) in SOPT and PBE functional used in SOPT
instead of B3LYP. The calculated TS structure agrees with that by GRRM,
and the activation energy in this case is 1.64 eV, which agrees well
with that of 1.51 eV by GRRM, and the difference is attributed to the
fact that GRRM used 6-31G as basis sets instead of 6-31G(d) in this
case. As for the failure of NEB-R with 5/7/9 images, it is most likely
originated from the rotation of CHOH molecule, because at least 5 out of
11 images are needed for molecular rotation in the MEP, as the images
7-11 shown in Fig.\ref{fig:Aldol-1-normal-images}. This case clearly
shows the efficiency of SOPT, which removes the rotational part of MEP,
and thus enables NEB with much fewer images.

\begin{figure}
\centering
\includegraphics{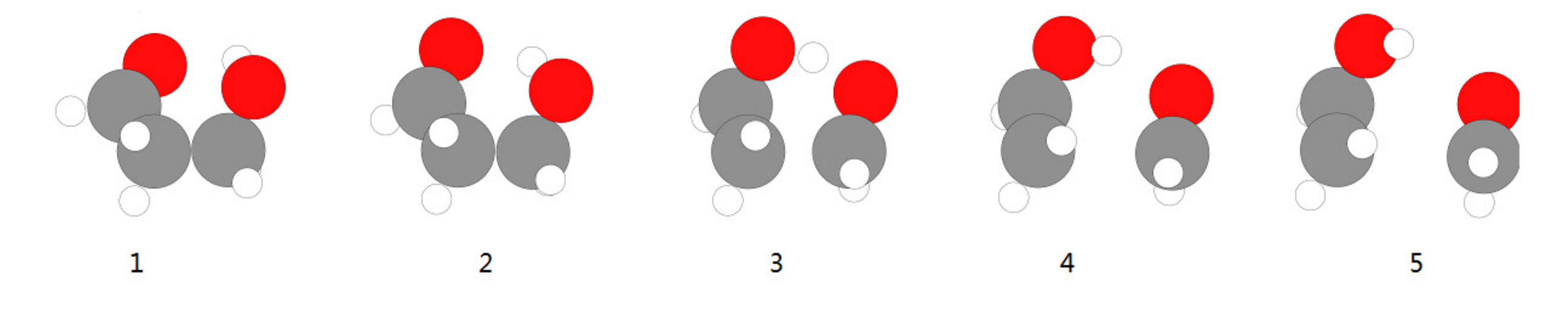}
\caption{The images in the Aldol-1 NEB5-SOPT MEP.
\label{fig:Aldol-1-SOPT-images}}
\end{figure}

\section{SUMMARY AND DISCUSSION}

We present the SOPT method as a robust and efficient method for
pre-optimizing model end structures used in the CoS methods like NEB,
instead of using the local minima. The SOPT method can generate more
relevant model end structures under the constraint of a given constant
RMSD distance between two ends to remove the unnecessary part(s)in the
MEP and thus accerlate CoS calculations. We demonstrate with two
examples that NEB-SOPT can reach convergence to the correct MEP and TS
with 60\% - 80\% less computational cost than NER-R, and NEB-SOPT allows
a small number of images, whereas NEB-R may need excessively large
numbers of both images and iterations to achieve the convergence for
difficult cases. The SOPT does not introduce extra computational cost in
place of regular geometry optimization, and enables lower level methods
like PBE or even LDA to deliver satisfactory results. A natural next
step is to extend the application of our method to heterogeneous
reactions.

\begin{acknowledgement}
This work was financially supported by the National Natural Science
Foundation of China (Grant Nos. 21590792, 91426302, and 21433005) to
J.L. and the Thousand Talents Plan for Young Scholars to H.X. The
calculations were performed using the supercomputers at the
Computational Chemistry Laboratory of Department of Chemistry under
Tsinghua Xuetang Talents Program. S.Z. would like to thank Chongyang
Zhao, Changsu Cao, and Xiaokun Zhao for their kind help.
\end{acknowledgement}


\bibliography{article}

\providecommand{\latin}[1]{#1}
\providecommand*\mcitethebibliography{\thebibliography}
\csname @ifundefined\endcsname{endmcitethebibliography}
  {\let\endmcitethebibliography\endthebibliography}{}
\begin{mcitethebibliography}{15}
\providecommand*\natexlab[1]{#1}
\providecommand*\mciteSetBstSublistMode[1]{}
\providecommand*\mciteSetBstMaxWidthForm[2]{}
\providecommand*\mciteBstWouldAddEndPuncttrue
  {\def\EndOfBibitem{\unskip.}}
\providecommand*\mciteBstWouldAddEndPunctfalse
  {\let\EndOfBibitem\relax}
\providecommand*\mciteSetBstMidEndSepPunct[3]{}
\providecommand*\mciteSetBstSublistLabelBeginEnd[3]{}
\providecommand*\EndOfBibitem{}
\mciteSetBstSublistMode{f}
\mciteSetBstMaxWidthForm{subitem}{(\alph{mcitesubitemcount})}
\mciteSetBstSublistLabelBeginEnd
  {\mcitemaxwidthsubitemform\space}
  {\relax}
  {\relax}

\bibitem[Henkelman(2000)]{Henkelman2000Improved}
Henkelman,~G. Improved tangent estimate in the nudged elastic band method for
  finding minimum energy paths and saddle points. \emph{Journal of Chemical
  Physics} \textbf{2000}, \emph{113}, 9978--9985\relax
\mciteBstWouldAddEndPuncttrue
\mciteSetBstMidEndSepPunct{\mcitedefaultmidpunct}
{\mcitedefaultendpunct}{\mcitedefaultseppunct}\relax
\EndOfBibitem
\bibitem[JÓNSSON \latin{et~al.}()JÓNSSON, MILLS, and JACOBSEN]{NEB-book}
JÓNSSON,~H.; MILLS,~G.; JACOBSEN,~K.~W. \emph{Classical and Quantum Dynamics
  in Condensed Phase Simulations}; pp 385--404\relax
\mciteBstWouldAddEndPuncttrue
\mciteSetBstMidEndSepPunct{\mcitedefaultmidpunct}
{\mcitedefaultendpunct}{\mcitedefaultseppunct}\relax
\EndOfBibitem
\bibitem[Sheppard \latin{et~al.}(2012)Sheppard, Xiao, Chemelewski, Johnson, and
  Henkelman]{NEB2012}
Sheppard,~D.; Xiao,~P.; Chemelewski,~W.; Johnson,~D.~D.; Henkelman,~G. A
  generalized solid-state nudged elastic band method. \emph{The Journal of
  Chemical Physics} \textbf{2012}, \emph{136}, 074103\relax
\mciteBstWouldAddEndPuncttrue
\mciteSetBstMidEndSepPunct{\mcitedefaultmidpunct}
{\mcitedefaultendpunct}{\mcitedefaultseppunct}\relax
\EndOfBibitem
\bibitem[Sheppard \latin{et~al.}(2008)Sheppard, Terrell, and
  Henkelman]{Sheppard2008Optimization}
Sheppard,~D.; Terrell,~R.; Henkelman,~G. Optimization methods for finding
  minimum energy paths. \emph{Journal of Chemical Physics} \textbf{2008},
  \emph{128}, 385--404\relax
\mciteBstWouldAddEndPuncttrue
\mciteSetBstMidEndSepPunct{\mcitedefaultmidpunct}
{\mcitedefaultendpunct}{\mcitedefaultseppunct}\relax
\EndOfBibitem
\bibitem[Sheppard and Henkelman(2011)Sheppard, and
  Henkelman]{Sheppard2011Paths}
Sheppard,~D.; Henkelman,~G. Paths to which the nudged elastic band converges.
  \emph{Journal of Computational Chemistry} \textbf{2011}, \emph{32},
  1769--1771\relax
\mciteBstWouldAddEndPuncttrue
\mciteSetBstMidEndSepPunct{\mcitedefaultmidpunct}
{\mcitedefaultendpunct}{\mcitedefaultseppunct}\relax
\EndOfBibitem
\bibitem[Sheppard \latin{et~al.}(2012)Sheppard, Xiao, Chemelewski, Johnson, and
  Henkelman]{Sheppard2012A}
Sheppard,~D.; Xiao,~P.; Chemelewski,~W.; Johnson,~D.~D.; Henkelman,~G. A
  generalized solid-state nudged elastic band method. \emph{Journal of Chemical
  Physics} \textbf{2012}, \emph{136}, 385--404\relax
\mciteBstWouldAddEndPuncttrue
\mciteSetBstMidEndSepPunct{\mcitedefaultmidpunct}
{\mcitedefaultendpunct}{\mcitedefaultseppunct}\relax
\EndOfBibitem
\bibitem[Smidstrup \latin{et~al.}(2014)Smidstrup, Pedersen, Stokbro, and
  Jónsson]{IDPP}
Smidstrup,~S.; Pedersen,~A.; Stokbro,~K.; Jónsson,~H. Improved initial guess
  for minimum energy path calculations. \emph{The Journal of Chemical Physics}
  \textbf{2014}, \emph{140}, 214106\relax
\mciteBstWouldAddEndPuncttrue
\mciteSetBstMidEndSepPunct{\mcitedefaultmidpunct}
{\mcitedefaultendpunct}{\mcitedefaultseppunct}\relax
\EndOfBibitem
\bibitem[Henkelman \latin{et~al.}(2000)Henkelman, Uberuaga, and
  Jónsson]{CINEB}
Henkelman,~G.; Uberuaga,~B.~P.; Jónsson,~H. A climbing image nudged elastic
  band method for finding saddle points and minimum energy paths. \emph{Journal
  of Chemical Physics} \textbf{2000}, \emph{113}, 9901--9904\relax
\mciteBstWouldAddEndPuncttrue
\mciteSetBstMidEndSepPunct{\mcitedefaultmidpunct}
{\mcitedefaultendpunct}{\mcitedefaultseppunct}\relax
\EndOfBibitem
\bibitem[Melander \latin{et~al.}(2015)Melander, Laasonen, and
  Jónsson]{dof-quaternion}
Melander,~M.; Laasonen,~K.; Jónsson,~H. Removing External Degrees of Freedom
  from Transition-State Search Methods using Quaternions. \emph{Journal of
  Chemical Theory and Computation} \textbf{2015}, \emph{11}, 1055--1062, PMID:
  26579757\relax
\mciteBstWouldAddEndPuncttrue
\mciteSetBstMidEndSepPunct{\mcitedefaultmidpunct}
{\mcitedefaultendpunct}{\mcitedefaultseppunct}\relax
\EndOfBibitem
\bibitem[Abashkin and Russo(1994)Abashkin, and Russo]{Abashkin1994Transition}
Abashkin,~Y.; Russo,~N. Transition state structures and reaction profiles from
  constrained optimization procedure. Implementation in the framework of
  density functional theory. \emph{Journal of Chemical Physics} \textbf{1994},
  \emph{100}, 4477--4483\relax
\mciteBstWouldAddEndPuncttrue
\mciteSetBstMidEndSepPunct{\mcitedefaultmidpunct}
{\mcitedefaultendpunct}{\mcitedefaultseppunct}\relax
\EndOfBibitem
\bibitem[Frisch \latin{et~al.}()Frisch, Trucks, Schlegel, Scuseria, Robb,
  Cheeseman, Scalmani, Barone, Mennucci, Petersson, Nakatsuji, Caricato, Li,
  Hratchian, Izmaylov, Bloino, Zheng, Sonnenberg, Hada, Ehara, Toyota, Fukuda,
  Hasegawa, Ishida, Nakajima, Honda, Kitao, Nakai, Vreven, Montgomery, Peralta,
  Ogliaro, Bearpark, Heyd, Brothers, Kudin, Staroverov, Kobayashi, Normand,
  Raghavachari, Rendell, Burant, Iyengar, Tomasi, Cossi, Rega, Millam, Klene,
  Knox, Cross, Bakken, Adamo, Jaramillo, Gomperts, Stratmann, Yazyev, Austin,
  Cammi, Pomelli, Ochterski, Martin, Morokuma, Zakrzewski, Voth, Salvador,
  Dannenberg, Dapprich, Daniels, Farkas, Foresman, Ortiz, Cioslowski, and
  Fox"]{g09}
Frisch,~M.~J.; Trucks,~G.~W.; Schlegel,~H.~B.; Scuseria,~G.~E.; Robb,~M.~A.;
  Cheeseman,~J.~R.; Scalmani,~G.; Barone,~V.; Mennucci,~B.; Petersson,~G.~A.;
  Nakatsuji,~H.; Caricato,~M.; Li,~X.; Hratchian,~H.~P.; Izmaylov,~A.~F.;
  Bloino,~J.; Zheng,~G.; Sonnenberg,~J.~L.; Hada,~M.; Ehara,~M.; Toyota,~K.;
  Fukuda,~R.; Hasegawa,~J.; Ishida,~M.; Nakajima,~T.; Honda,~Y.; Kitao,~O.;
  Nakai,~H.; Vreven,~T.; Montgomery,~J.~A.,~{Jr.}; Peralta,~J.~E.; Ogliaro,~F.;
  Bearpark,~M.; Heyd,~J.~J.; Brothers,~E.; Kudin,~K.~N.; Staroverov,~V.~N.;
  Kobayashi,~R.; Normand,~J.; Raghavachari,~K.; Rendell,~A.; Burant,~J.~C.;
  Iyengar,~S.~S.; Tomasi,~J.; Cossi,~M.; Rega,~N.; Millam,~J.~M.; Klene,~M.;
  Knox,~J.~E.; Cross,~J.~B.; Bakken,~V.; Adamo,~C.; Jaramillo,~J.;
  Gomperts,~R.; Stratmann,~R.~E.; Yazyev,~O.; Austin,~A.~J.; Cammi,~R.;
  Pomelli,~C.; Ochterski,~J.~W.; Martin,~R.~L.; Morokuma,~K.;
  Zakrzewski,~V.~G.; Voth,~G.~A.; Salvador,~P.; Dannenberg,~J.~J.;
  Dapprich,~S.; Daniels,~A.~D.; Farkas,~O.; Foresman,~J.~B.; Ortiz,~J.~V.;
  Cioslowski,~J.; Fox",~D.~J. Gaussian 09 {R}evision {E}.01. Gaussian Inc.
  Wallingford CT 2009\relax
\mciteBstWouldAddEndPuncttrue
\mciteSetBstMidEndSepPunct{\mcitedefaultmidpunct}
{\mcitedefaultendpunct}{\mcitedefaultseppunct}\relax
\EndOfBibitem
\bibitem[Larsen \latin{et~al.}(2017)Larsen, Mortensen, Blomqvist, Castelli,
  Christensen, Dułak, Friis, Groves, Hammer, Hargus, Hermes, Jennings, Jensen,
  Kermode, Kitchin, Kolsbjerg, Kubal, Kaasbjerg, Lysgaard, Maronsson, Maxson,
  Olsen, Pastewka, Peterson, Rostgaard, Schiøtz, Schütt, Strange, Thygesen,
  Vegge, Vilhelmsen, Walter, Zeng, and Jacobsen]{ase}
Larsen,~A.~H.; Mortensen,~J.~J.; Blomqvist,~J.; Castelli,~I.~E.;
  Christensen,~R.; Dułak,~M.; Friis,~J.; Groves,~M.~N.; Hammer,~B.;
  Hargus,~C.; Hermes,~E.~D.; Jennings,~P.~C.; Jensen,~P.~B.; Kermode,~J.;
  Kitchin,~J.~R.; Kolsbjerg,~E.~L.; Kubal,~J.; Kaasbjerg,~K.; Lysgaard,~S.;
  Maronsson,~J.~B.; Maxson,~T.; Olsen,~T.; Pastewka,~L.; Peterson,~A.;
  Rostgaard,~C.; Schiøtz,~J.; Schütt,~O.; Strange,~M.; Thygesen,~K.~S.;
  Vegge,~T.; Vilhelmsen,~L.; Walter,~M.; Zeng,~Z.; Jacobsen,~K.~W. The atomic
  simulation environment—a Python library for working with atoms.
  \emph{Journal of Physics: Condensed Matter} \textbf{2017}, \emph{29},
  273002\relax
\mciteBstWouldAddEndPuncttrue
\mciteSetBstMidEndSepPunct{\mcitedefaultmidpunct}
{\mcitedefaultendpunct}{\mcitedefaultseppunct}\relax
\EndOfBibitem
\bibitem[Bahn and Jacobsen(2002)Bahn, and Jacobsen]{ase-old}
Bahn,~S.~R.; Jacobsen,~K.~W. An object-oriented scripting interface to a legacy
  electronic structure code. \emph{Comput. Sci. Eng.} \textbf{2002}, \emph{4},
  56--66\relax
\mciteBstWouldAddEndPuncttrue
\mciteSetBstMidEndSepPunct{\mcitedefaultmidpunct}
{\mcitedefaultendpunct}{\mcitedefaultseppunct}\relax
\EndOfBibitem
\bibitem[Maeda \latin{et~al.}(2013)Maeda, Ohno, and Morokuma]{GRRM}
Maeda,~S.; Ohno,~K.; Morokuma,~K. Systematic exploration of the mechanism of
  chemical reactions: the global reaction route mapping (GRRM) strategy using
  the ADDF and AFIR methods. \emph{Phys. Chem. Chem. Phys.} \textbf{2013},
  \emph{15}, 3683--3701\relax
\mciteBstWouldAddEndPuncttrue
\mciteSetBstMidEndSepPunct{\mcitedefaultmidpunct}
{\mcitedefaultendpunct}{\mcitedefaultseppunct}\relax
\EndOfBibitem
\end{mcitethebibliography}

%

\end{document}